\documentclass[12pt]{aastex}

\usepackage{amsmath}
\shorttitle{Water formation via ozone hydrogenation} 
\shortauthors{Mokrane et al.}

\begin{document}

\title{Experimental evidence for water formation via ozone hydrogenation on
dust grains at 10~K}

\author{
H. Mokrane \altaffilmark{1}, H. Chaabouni \altaffilmark{1}, M.
Accolla \altaffilmark{1,2},
 E. Congiu \altaffilmark{1},
 F. Dulieu\altaffilmark{1},
 M. Chehrouri \altaffilmark{1,3},
 J.~L.~Lemaire\altaffilmark{1} }

\altaffiltext{1}{LERMA/LAMAp, UMR 8112 du CNRS, de l'Observatoire de
Paris et de l'Universit\'e de Cergy Pontoise, 5 mail Gay Lussac,
95000 Cergy Pontoise, France}
\altaffiltext{2}{Dipartimento di
Metodologie Fisiche e Chimiche, Universit\`a di Catania, Viale A.
Doria 6,95125 Catania, Sicily, Italy}
\altaffiltext{3}{Universit\'e
de Sa\"{\i}da, BP 138 Enaser Sa\"{\i}da 2002, Alg\'erie}

\begin{abstract}
The formation of water molecules from the reaction between ozone
(O$_3$) and D-atoms is studied experimentally for the first time.
Ozone is deposited on non-porous amorphous solid water ice (H$_2$O),
and D-atoms are then sent onto the sample held at 10~K. HDO
molecules are detected during the desorption of the whole substrate
where isotope mixing takes place, indicating that water synthesis
has occurred. The efficiency of water formation via hydrogenation of
ozone is of the same order of magnitude of that found for reactions
involving O atoms or O$_2$ molecules and exhibits no apparent
activation barrier. These experiments validate the assumption made
by models using ozone as one of the precursors of water formation
via solid-state chemistry on interstellar dust grains.

\end{abstract}


\section{Introduction}

Solid water is a very abundant material. Indeed, it is believed to
be the most abundant condensed material in the Universe, thanks to
its propensity for remaining as a deposit on interstellar dust
particles in dense clouds. Amorphous solid water ice, together with
a wealth of other condensed species, is widely observed
spectroscopically via infra-red absorption lines present in the
spectrum of a field star or of the embedded object in star forming
regions \citep{gibb00,boogert04}. Icy mantles covering dust grains
are reputed to form after direct freeze out of gas phase species and
after surface reactions of atoms and radicals.

Although water ice is by far the most abundant species in icy
mantles, the chemical origin of water molecules in interstellar icy
mantles is still uncertain. Water molecule formation in the gas
phase is not efficient enough to reproduce the observed abundances
in dark clouds, especially in its solid form \citep{pariseb,
ceccarelli}. Therefore, water ice is likely to form directly on the
cold interstellar grains and not as a condensate after formation in
the gas phase.

A reaction scheme including water formation on grain surfaces was
proposed some years ago by \citet{tielensa}. They suggested that
H$_{2}$O formation would be initiated by H-atoms reacting with O,
O$_{2}$ and O$_{3}$, although the O$_{3}$~+~H pathway was considered
to be the most effective and O$_{2}$ would play a mere catalytic
role.

Nonetheless, O$_2$ and O$_3$ molecules have not so far been detected
in interstellar ices on dust grains, although their presence is
presumed.  Ozone non-detection may be partly explained because its
strongest $\nu_3$ (O-O) vibrational transition band at 1040
cm$^{-1}$ (9.6 $\mu$m) is swamped by the strong infrared absorption
band of silicates at 10~$\mu$m which has been observed by ISO
(Infrared Space Observatory)  in the 2.4-25 $\mu$m region
\citep[e.g., spectrum toward the protostar W33A, resolution
$\sim$~500-1000,][]{gibb00}.

Astrochemical models of dark clouds predict that condensed
oxygen is likely to be a major component of apolar dust grains
within interstellar clouds at 10~K. Recently, the O~+~H pathway
\citep{dulieu09} and the O$_2$~+~H pathway
\citep{miyauchi,ioppolo,matar} have been explored experimentally.
All these experiments show that these reactions lead to water
formation, even though the efficiencies of intermediate reactions
are still under debate. Besides O and O$_2$, ozone too can be
present in icy mantles chemically processed by irradiation from
Galactic cosmic-ray particles and internal secondary UV field
\citep{tielensa,mathis83,greenberg84,strazzulla91}. Several
laboratory works confirm the efficient production of O$_3$ following
energetic electron irradiation of pure oxygen ices
\citep{bennett05,sivaraman07}, and UV, proton and ion irradiation of
ice mixtures simulating realistic analogues of interstellar ices
\citep{ehrenfreund92,strazzulla97,cooper08}. In solid phase, O$_2$
and CO$_2$ are easily dissociated by UV radiations ($\lambda~<
~240$~nm) and cosmic rays~(c.r.) producing atomic O in the
electronic ground state ($^3$P) and in the first electronically
excited ($^1$D) state \citep{sivaraman07}:
\begin{equation}\label{1}
    \mathrm{O}_{2} \xrightarrow{h\nu
    ,c.r.} \mathrm{O} + \mathrm{O}
\end{equation}
\begin{equation}\label{2}
    \mathrm{CO}_{2} \xrightarrow{h\nu
    ,c.r.} \mathrm{CO} + \mathrm{O}
\end{equation}
In the interstellar grain mantles, the O($^3$P) atoms in the ground
state, which are in thermal equilibrium with the ice surface, react
with O$_2$ molecules to form O$_3$ via the following reaction
pathway:
\begin{equation}\label{3}
    \mathrm{O} + \mathrm{O}_{2} \longrightarrow
    \mathrm{O}_{3}
\end{equation}
The formation of ozone via equation (3) is exothermic
($\Delta_fH\degr=-1.48$~eV).

While the main ozone destruction reactions are:
\begin{equation}\label{4}
    \mathrm{O}_{3} + \mathrm{H} \longrightarrow \mathrm{OH} +
    \mathrm{O}_{2}
\end{equation}
\begin{equation}\label{5}
    \mathrm{O}_{3} \xrightarrow{h\nu
    ,c.r.} \mathrm{O}_{2} + \mathrm{O}
\end{equation}
In addition, in dark clouds hydrogen is mainly in its molecular form
so H-atoms are a rather rare reactant with H/H${_2}$ $\sim 10^{-3}$
\citep{li03}. The number density of H-atoms is mostly governed by
the destruction of H${_2}$ due to cosmic rays. This value, whatever
the density of the cloud, is about 1~cm$^{-3}$. The O/H${_2}$ ratio
remains approximately constant (10$^{-4}$), thus the number of
atomic O, unlike H, is proportional to the density of the cloud.
See, for example, Table~1 of \citet{caselli};  for a cloud density
of 10$^4$~cm$^{-3}$ the H/O ratio is $\sim 1/0.75$ while for a
denser cloud with a density of 10$^5$~cm$^{-3}$ the H/O ratio is
$\sim 1/7$. Therefore, for very dense clouds, O is the most abundant
species in atomic form and it can accrete on grains and subsequently
form O$_2$ and O$_3$.

 In their model, using a Monte Carlo approach, \citet{cuppen} showed
that under dense cloud conditions the (O$_3$~+~H) pathway is the
most efficient route leading to the formation of H$_2$O. Finally, it
should be noted that, in the diverse models, some barriers can be
raised in order to increase one of the competitive reactions
\citep{lee,caselli,parise04}. Yet, such barriers can only be
experimentally probed, because extrapolations from the gaseous data
are quite uncertain.

This Letter is the first report of water formation via ozone
hydrogenation in the series of recent experiments that aim to
investigate water synthesis in the ISM. We present the evidence for
the formation of water molecules from the reaction of O$_3$ with
D-atoms on non-porous amorphous solid water ice under conditions
relevant to the dense interstellar clouds.

\section{Experimental procedures}
Our experiments were performed using the FORMOLISM set-up developed
to study the reaction and the interaction of atoms and molecules on
surfaces simulating dust grains under interstellar conditions. The
apparatus is composed of an ultrahigh-vacuum chamber with a base
operating pressure of around 10$^{-10}$ mbar. The sample holder, a
copper cylinder block, is attached to the cold finger of a
closed-cycle He cryostat. It can be cooled to 8~K. The temperature
of the sample is measured using a calibrated silicon diode and a
thermocouple (Cromel-AuFe) clamped on the back of the sample holder.
The copper surface is covered with 100 layers of non-porous
amorphous solid water (np-ASW) by condensation of water vapor on the
cold sample maintained at 120~K. Water is sprayed through a
microchannel array doser located 2~cm in front of the surface. This
ice is amorphous but non porous \citep{kimmel}. The water ice
substrate is then cooled to 10~K. More details can be obtained in
\citet{amiaud}.

Ozone can be synthesized by two methods. One is an ex-situ method:
O$_3$ is prepared from gaseous O$_2$ introduced in a glass bottle at
a pressure of about 30~mbar. The O$_2$ gas is then excited by a
radiofrequency electric discharge ($\sim$~2~MHz) through a copper
coil placed around the O$_2$ container. The blue ozone gas produced
by the recombination of O$_2$ and O atoms during the discharge is
condensed on the bottom wall plunged in liquid nitrogen (77~K).
Residual O$_2$ is then removed by primary pumping. The glass bottle
is then mounted onto the main set-up. Ozone is introduced into the
vacuum chamber via a triply differentially pumped beam line aimed at
the cold surface covered with H$_2$O ice. Due to the metallic parts
of the inlet system on which O$_3$ in part dissociates, the beam is
composed of 70~\% ozone and 30~\% O$_2$. However, O$_2$ is easily
removed from the surface when the sample is held at 50~K. By varying
the exposure time of the ozone beam, it is possible to vary the
initial coverage of the surface. The O$_3$ flux was previously
calibrated using temperature programmed desorption (TPD) by
determining the exposure time of a pure O$_2$ beam required to
saturate the O$_2$ monolayer (1~monolayer = 10$^{15}$
molecules~cm$^{-2}$) on np-ASW ice. In the experiments described
below we have deposited a maximum of half a monolayer of ozone under
the same beam conditions used for calibrating the O$_2$ flux. Owing
to the non-pure O$_3$ beam, however, ozone exposures had a relative
accuracy of $\sim$~20~\% and an absolute accuracy of $\sim$~30~\%.
Nevertheless, we did not see any significant effect due to the
initial coverage (between 0.1 and 0.5 monolayers~$\pm~30~\%$).

 An alternative technique of ozone production was also used, namely
by depositing an atomic O on top of an O$_2$ layer pre-adsorbed on
the surface and then producing ozone in-situ. The disadvantage of
this technique is that the control of the initial amount of ozone
have to be checked after each exposure. Whichever technique was used
to deposit ozone onto the surface, we obtained the same final
results.

After depositing ozone, an atomic D-beam aimed at the surface of the
sample irradiates the O$_3$-H$_2$O film maintained at 10~K. The
D-atoms are produced by the dissociation of D$_2$ molecules in a
quartz tube using a Surfatron microwave discharge at 2.45~GHz. The
rate of dissociation is $\sim$ 60~\%. After the irradiation of $\sim
5 \times 10^{15}$ D-atoms~cm$^{-2}$, a TPD is performed up to 180~K
so that all the water ice desorbs. The TPD heating rate is 10 K/min
from 10 to 180~K. The desorbed species are detected using a
quadruple mass spectrometer (QMS).

\section{Results and discussion}

\begin{figure}
\includegraphics[width=8cm]{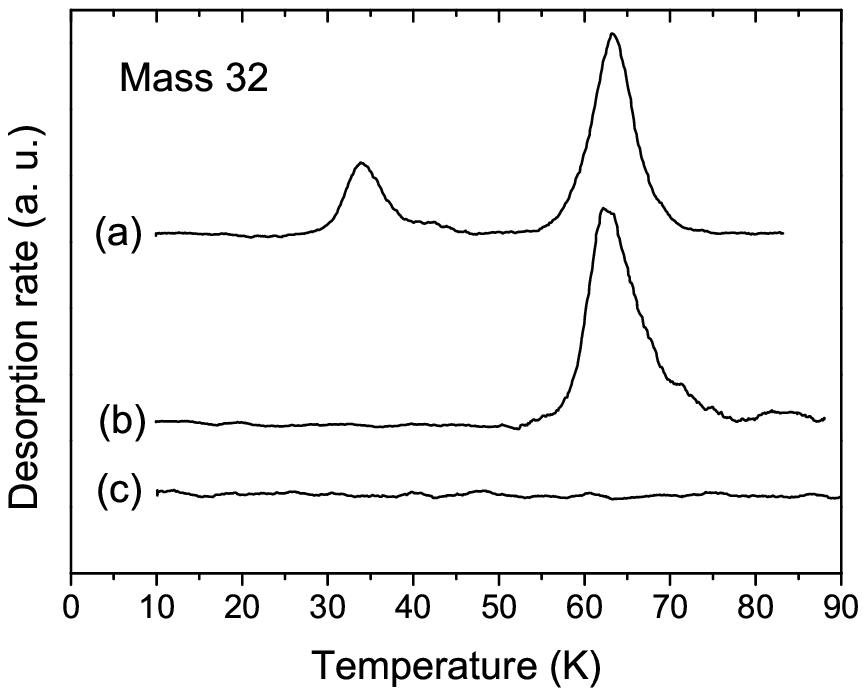}\centering \\
\caption{TPD spectra of mass~32 obtained after depositing
 $\sim 5\times10^{14}$ molecules~cm$^{-2}$ of O$_3$ on np-ASW: (a) O$_3$ deposited
at 10~K, (b) O$_3$ deposited at 50~K and cooled to 10~K, (c) O$_3$
deposited at 50~K, cooled to 10~K and irradiated with
$5\times10^{15}$ cm$^{-2}$ of D-atoms.} \label{Fig1}
\end{figure}

The spectrum (a) in Figure~\ref{Fig1} shows the TPD curve of mass~32
obtained after the exposure of half a layer of O$_3$ on a np-ASW
substrate held at 10~K (ex-situ synthesis). We observe two peaks:
the desorption peak at low temperature ($\sim$~30~K) represents the
desorption of O$_2$ molecules. These molecules have been deposited
by the O$_2$ fraction in the O$_3$ beam, probably due to ozone
destruction on the metallic walls of the gas inlet system. The
high-temperature peak at $\sim$~60~K is the desorption of ozone.
Mass 48 of O$_3$ is also simultaneously detected in smaller amount
and is always proportional to the signal at mass 32. In fact, for
the most part O$_2^+$ fragments are detected, as a result of the
O$_3$ cracking in the head of the QMS. Also, the desorption of ozone
from ASW surfaces is known to occur at around 60~K
\citep{chaabouni00,borget}. Although this may be seen as an indirect
measure, mass~32 gives information about O$_2$ (below 50~K) and
O$_3$ (above 50~K) desorption. As shown in the spectrum (a) of
Fig.~1, O$_3$ desorption is about three times larger than that of
O$_2$. This indicates that the ozone fraction deposited with the
ex-situ method is $\sim$~70~\%, the rest being O$_2$ molecules. In
order to deposit only ozone molecules on the cold surface, we
performed the exposure step with the water ice substrate held at
50~K. Thus O$_2$ molecules desorbing at 30~K were eliminated. Then
we cooled the sample to 10~K before proceeding with the TPD. As
shown in the spectrum (b) of Fig.~1, the TPD peak observed comes
from O$_3$ desorbing at 60~K. By using these procedures, we are sure
that at 10~K only reactions involving ozone and D-atoms will occur.

For this work, we performed a set of experiments with various D-atom
exposures. In the first one, D-atoms were sent onto the ozone-free
water ice surface to check that no reaction occurred between D and
the H$_2$O substrate to produce deuterated water molecules. This
result is displayed in the mass~19 spectrum (b) of Fig.~2. There is
neither HDO nor D$_2$O desorption peak when D-atoms alone are
deposited on the ASW ice substrate at 10~K as was previously noticed
by \citet{nagaoka} and \citet{dulieu09}.

In a second experiment we exposed the np-ASW sample held at 10~K to
D-atoms before ozone deposition. In this experiment ozone desorbs as
if no atoms had been sent to the surface. This is because D-atoms
form promptly D$_2$ on non-porous water ice \citep[e.g.,][]{congiu}
and apparently because ozone does not react with D$_2$ (see below).
This proves that hydrogenation reactions take place at around 10~K
and not at higher temperatures since D-atoms react promptly at 10~K
and furthermore they will have completely desorbed above 13~K
\citep{amiaud07}.

In another experiment, D-atoms were sent onto the np-ASW surface
previously exposed to 0.5 monolayers of O$_3$. In the spectrum (c)
of Figure~1 we firstly observe the disappearance of the O$_3$ peak.
This suggests that all the ozone molecules have reacted. We also
note that the reaction D~+~O$_3$ is very efficient, because the
number of D-atoms sent to the surface is only 10~times the number of
O$_3$ on the surface. Consequently, we can conclude that the
D~+~O$_3$ reaction
--- which is in competition with the D~+~D reaction in our
experiment --- proceeds on the ice surface with a small energy
barrier. An estimation of the reaction rate can be made, for
example, with the formula used by \citet{cuppen} to simulate the
surface reactions in their model:
\begin{equation}\label{6}
   R_{react}=\nu_{r}exp~(-\frac{2a}{h}\sqrt{2E_\mathrm{a}\mu}),
\end{equation}
where $\nu_{r}$ is the attempt frequency for reaction, $a$ the width
of the barrier, $\mu$ the reduced mass and $E_\mathrm{a}$ the
activation barrier. By taking $E_\mathrm{a}$/k$_\mathrm{B}$~=~450~K
for the D~+~O$_3$ reaction \citep{lee,cuppen}, we found that this
reaction is very rapid ($\sim$~1~ps) if compared with the diffusion
time of D-atoms on an amorphous water ice
\citep[$\tau_\mathrm{diff}$~=~$\sim$~10~ms~on porous ASW
ice,][]{matar}.

In passing, we would like to point out that, over the last years, we
have performed experiments on water formation in space starting from
O, O$_2$ and now O$_3$. Even though we cannot infer absolute
reaction rates which are very hard to estimate we can attempt to
make relative comparisons.
It should be noted that the destruction
rate of ozone following D exposures is as important as the one for
O~+~D and D~+~O$_3$ reactions within the error bar of our
experimental data. Hence we conclude that the D~+~O$_3$ reaction
proceeds without a measurable activation barrier in the solid phase.

\begin{figure}
\includegraphics[width=8cm]{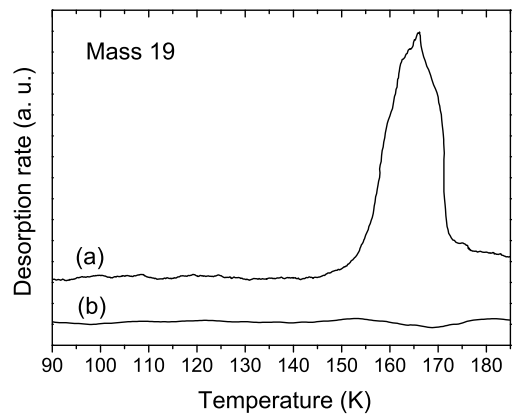}\centering \\
\caption{TPD spectra of mass~19 (HDO) obtained after: (a) deposition
of $5\times10^{15}$~cm$^{-2}$~D-atoms on the np-ASW substrate at
10~K previously exposed to $5\times10^{14}$ molecules~cm$^{-2}$ of
O$_3$, (b) deposition of $2\times10^{16}$ cm$^{-2}$ D-atoms on the
ozone-free water ice surface at 10~K .} \label{Fig2}
\end{figure}

Obtained subsequently to the D~+~O$_3$ experiment, the spectrum (a)
of Figure~\ref{Fig2} shows a desorption peak of mass~19 at 160~K
attributed to the singly deuterated water (HDO). This clearly
indicates that the reaction of D-atoms with O$_3$ molecules on the
ice surface is efficient and produces deuterated water molecules
detected upon desorption of the whole ice film. We also notice that
nearly no desorption peak of D$_2$O (mass~20) is detected. This is
due to thermally activated H/D exchanges during the phase transition
of the water ice \citep{smith97}. Therefore, even if D$_2$O is
actually formed on the surface, mostly water molecules in the form
of HDO are likely to desorb and be detected.

The formation of deuterated water molecules at T$_s$~= 10~K from
D-atoms and O$_3$ molecules may then follow the successive reaction
pathways:
\begin{equation}\label{7}
\mathrm{O}_{3} + \mathrm{D} \longrightarrow \mathrm{OD} + \mathrm{O}_{2}
\end{equation}
\begin{equation}\label{8}
\mathrm{OD} + \mathrm{D} \longrightarrow \mathrm{{D}_{2}{O}}
\end{equation}
\begin{equation}\label{9}
\mathrm{O}_{2} + \mathrm{D} \longrightarrow \mathrm{DO}_{2}
\end{equation}
\begin{equation}\label{10}
\mathrm{DO}_{2} + \mathrm{D} \longrightarrow \mathrm{{D}_{2}{O}_{2}}
\end{equation}
\begin{equation}\label{11}
\mathrm{{D}_{2}{O}_{2}} + \mathrm{D} \longrightarrow
\mathrm{{D}_{2}{O}} + \mathrm{D}
\end{equation}
Whereas at T$_s$ greater than 130~K, as mentioned above, the
following isotope exchange reaction is active:
\begin{equation}\label{12}
\mathrm{{H}_{2}{O}} + \mathrm{{D}_{2}{O}} \longrightarrow
\mathrm{{2}~{HDO}}
\end{equation}

Of course, the reversal of reaction (12) is also possible, but in
our case it is statistically negligible.

As for what concerns the intermediate products of the reaction
scheme, we performed again the D~+~O$_3$ experiment but with a low
exposure of D-atoms (0.01 monolayers) on 0.15 monolayers of O$_3$,
thus with a dose of D insufficient to go on with the hydrogenation
reactions leading to water molecules. This experiment shows that
O$_3$ molecules are not thoroughly consumed and also that O$_2$
molecules appear on the surface indicating that reaction (7)
proceeds at 10~K and that the formation of D$_2$O can proceed as
well from the intermediate reaction (9).

The studied mechanism for water formation from the hydrogenation of
ozone might be an explanation for the non- detection of O$_3$ in the
interstellar ices. In fact, ozone reacts rapidly with hydrogen atoms
to form water molecules on a water ice surface at 10 K and under
conditions similar to those found in dense interstellar clouds.
Likewise, the lack of a clear detection of O$_2$ in interstellar
ices may be explained by the high reactivity between O$_2$ and
H-atoms reported in Miyauchi et al. 2008, Ioppolo et al. 2008, and
Matar et al. 2008.

In addition, the OH~+~H$_2$ reaction could also be a key surface
reaction in dark clouds, because H$_2$ (and D$_2$) is present on the
surface of the grains in large amounts \citep{kristensen09}. In our
experiments too, there is a large amount of molecular deuterium
because of the incomplete dissociation of D$_2$ in the D beam. At
the moment it is not possible to us to probe explicitly the
OH~+~H$_2$ reaction, but we can investigate the role of D$_2$, by
testing the O$_3$~+~ D$_2$ reaction. In order to do that, we have
deposited a very large amount of D$_2$ ($\sim$~10$^{17}$~cm$^{-2}$)
on np-ASW ice held at 10 K and previously dosed with O$_3$. In this
experiment, we did not detect any HDO, even when the surface was
held at 50~K during D$_2$ deposition in order to overcome a possible
activation barrier. Therefore, we can fairly conclude that the
D$_2$~+~O$_3$ reaction does not proceed under our experimental
conditions.

Unlike the O$_3$~+~D$_2$ reaction, O$_3$~+~D shows no barrier under
our experimental conditions and this result shows clearly that ozone
is a likely precursor to water formation on a water ice layer at
10~K as proposed by \citet{cuppen}. This pathway is to be preferred
in environments where the number of O atoms exceeds the number of
H~atoms, which is supposed to be the case in dark clouds with a
density higher than 10$^4$ cm$^{-3}$ \citep{caselli}. Unfortunately,
our present data do not allow us to disentangle the different
intermediate reactions that lead to the formation of water. In order
to investigate in more detail the kinetics of water formation at
10~K via ozone, we are developing an additional IR diagnostic on
FORMOLISM. This will allow us to see the appearance of some of the
intermediate reactants and to provide accurate quantitative results.

\section{Conclusions}
In this paper we have shown that it is possible to form efficiently
water molecules from the reaction of O$_3$ with hydrogen atoms on
non-porous water ice. This reaction exhibits no activation barrier
at 10~K, and is about as efficient as the pathways having O or O$_2$
as precursors. This experimental result corroborates theoretical
studies that consider ozone as one of the major actors of water
formation in dark clouds.

\acknowledgments We acknowledge the support of the French PCMI
program funded by the CNRS, as well as the strong financial support
from the Conseil R\'egional d'Ile de France (SESAME program
I-07-597/R), the Agence Nationale de la Recherche (ref. 07-0129) and
the Conseil G\'en\'eral du Val d'Oise.

\email{Henda.Chaabouni@u-cergy.fr}.

\clearpage

\end{document}